\documentclass[usenatbib]{mn2e}
\usepackage{epsfig}

\def\beq{\begin{equation}}
\def\eeq{\end{equation}}
\def\mgal{M_{\rm gal}}
\def\rmerg{R_{\rm merg}}
\def\rmaj{R_{\rm major}}
\def\zres{z_{\rm res}}
\def\Msun{M_{\odot}}

\def\kms{{\rm km} \ {\rm s}^{-1}}
\def\cm3{{\rm cm^{-3}}}
\def\cm2{{\rm cm^{-2}}}

\title{Galaxy Merger Statistics and Inferred Bulge-to-Disk Ratios in
Cosmological SPH Simulations}

\author[Maller et al.]
{Ariyeh H. Maller$^{1,2}$, Neal Katz$^1$, Du\v{s}an Kere\v{s}$^1$, 
Romeel Dav\'e$^3$, David H. Weinberg$^4$\\
$^1$Astronomy Department, University of Massachusetts Amherst, 
710 N. Pleasant St., Amherst, MA 01003\\
$^2$Department of Physical and Biological Sciences, 
New York City College of Technology, 300 Jay St., Brooklyn, NY  11201\\
$^3$University of Arizona, Steward Observatory, 933 North Cherry Avenue, 
Tucson, AZ 85721\\
$^4$Department of Astronomy, Ohio State University, 140 West 18th Avenue, 
Columbus, OH 43210
}
\begin{document}

\maketitle

\begin{abstract}
We construct merger trees for galaxies identified in a cosmological 
hydrodynamical simulation and use them to characterize predicted merger rates
as a function of redshift, galaxy mass, and merger mass ratio.  At $z=0.3$,
we find a mean rate of 0.054 mergers per galaxy per Gyr above a 1:2 mass ratio
threshold for massive galaxies (baryonic mass above $6.4\times 10^{10}\Msun$),
but only 0.018 Gyr$^{-1}$ for lower mass galaxies.  The mass ratio distribution
is $\propto R_{\rm merg}^{-1.2}$ for the massive galaxy sample, so high mass
mergers dominate the total merger growth rate.  The predicted rates increase
rapidly with increasing redshift, and they agree reasonably well with 
observational estimates.  A substantial fraction of galaxies do not experience
any resolved mergers during the course of the simulation, and even for the 
high mass sample only 50\% of galaxies experience a greater than $1:4$ 
merger since $z=1$.  Typical galaxies thus have fairly quiescent merger 
histories.  

We assign bulge-to-disk ratios to simulated galaxies by assuming 
that mergers above a mass ratio threshold $R_{\rm major}$ convert stellar disks
into spheroids.  With $R_{\rm major}$ of $1:4$, we obtain a fairly good match 
to the observed dependence of early-type fraction on galaxy mass.  However, 
the predicted fraction of truly bulge-dominated systems ($f_{\rm bulge} > 0.8$)
is small, and producing a substantial population of bulge-dominated galaxies
may require a mechanism that shuts off gas accretion at late times and/or 
additional processes (besides major mergers) for producing bulges.
\end{abstract}


\section{Introduction}

It has long been suggested that the spheroidal components of
galaxies are a result of galaxy mergers (e.g., \citealt{tt:72}), 
and that a galaxy's merger history therefore plays a central role
in determining its Hubble type.  
Numerical simulations of disk galaxy mergers produce systems with
many of the properties of observed elliptical galaxies 
(e.g., \citealt{barn:88,hern:92,bh:96,nbh:99,spri:00,bb:00,cret:01,nb:03,
cox:04,sh:04}), including the telltale low surface brightness shells
and tidal features that are often revealed by deep imaging
(e.g., \citealt{malin:83,schw:90,dokkum:05}).
However, present simulations do not have the dynamic range needed to track 
the morphological evolution of individual galaxies while simultaneously 
modeling representative
cosmological volumes, so it is unclear whether the merger hypothesis
can explain the observed frequency and environment dependence of
Hubble types.  Semi-analytic models of galaxy formation have provided
the main avenue for progress on this question, using simplified
recipes to convert disk material in stellar bulges following mergers
\citep[e.g.][]{kwg:93,kauf:96b,baugh:96,sp:99,cole:00,bell:03,hatt:03,ny:04}.
These models have achieved reasonable success in reproducing observed
galaxy distributions, but they rest on an approximate treatment of
dynamical friction to compute galaxy merger rates from dark halo 
merger rates, in addition to the assumptions of the merger recipes
themselves.  One recent semi-analytic study that tracks ``galaxies'' 
as sub-halos in high resolution N-body simulations yields results
significantly different from those of other models \citep{kang:04}.
In addition, the effects of dynamical friction and tidal disruption 
on sub-halos will be altered by the presence of tightly
bound gas and stars within them.

In this paper, we compute the merger histories of galaxies formed
in a smoothed particle hydrodynamics (SPH) cosmological simulation.
While this simulation does not have the resolution required for
reliable morphological analysis, it incorporates all of the physics
that should have a major effect on merger rates, and we can apply
semi-analytic style recipes to compute resulting distributions
of Hubble types.  We can also address, at least in part, the related
issues of survival and angular momentum of galaxy disks.

Stellar disks are fragile, and one important question for cosmological
models is whether they predict a low enough merger rates to
preserve thin disks \citep{to:92}.
In addition, disk galaxies forming in cosmological SPH simulations
frequently lose angular momentum in late-time merger events, and
the simulated galaxies have, on average, less angular momentum than
observed disks \citep{ns:00}.  Recent simulations have
produced some examples of disk galaxies with realistic angular
momentum properties \citep{gover:04,rysh:04}, but
these results are strongly dependant on the galaxy's merger history.  
Therefore, without knowing the distribution of galaxy merger
histories we cannot ascertain how often such realistic disks are created.
It has also become clear that a halo's merging history is related 
to its angular momentum \citep{mds:02,vitv:02}, and 
the way that baryons trace that
merging history may play an important role in determining disk sizes 
\citep{md:02}.

Our results for the merger statistics of SPH galaxies complement
analytic and numerical calculations of dark halo merger statistics
\citep{kw:93,lc:93,lc:94,sk:99,slkd:00,wech:02}.
They significantly improve on the earlier study of \cite{mura:02}
by using a higher resolution simulation and, equally important,
by tracking merger histories of individual systems rather than
globally averaged merger and accretion rates.
We describe the simulation and our method of constructing
merger trees in \S\ref{sec:sims}.  We present merger statistics
in \S\ref{sec:prop}, and in \S\ref{sec:btd} we use the merger
histories to compute Hubble type distributions.
We summarize our results and discuss directions for
future work in \S\ref{sec:conc}.

\begin{figure} 
\centering 
\vspace{0pt} 
\epsfig{file=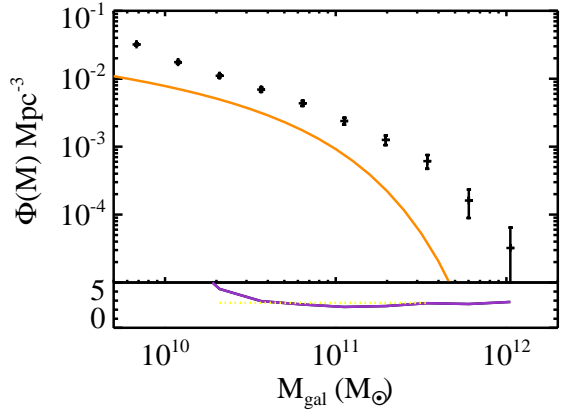,width=\linewidth} 
\vspace{0pt}
\caption{The upper panel shows the cumulative galaxy mass function at $z=0$ 
for our simulation and, for comparison, the Schechter function fit of 
\citet{bmkw:03}. The y-axis is in comoving units with $h=0.7$ as we use 
throughout this paper.  The simulations produce far too many low and high mass 
galaxies, while galaxies around the bend in the Schechter function 
are a factor of 2-3 too massive.  The bottom panel shows the factor 
by which the simulated galaxies masses should be divided in order to agree
with the observations. For the galaxy masses in which we 
are interested, $2 \times 10^{10} \Msun < \mgal < 6 \times 10^{11} \Msun$, 
the correction is roughly a factor of 2.75 as shown by the dotted line, and 
it does not depend strongly on galaxy mass.
}\label{fig:mf}
\end{figure}

\begin{table*}
\begin{center}
\begin{tabular}{cccccccccc}
\hline
Sample  & Number & Mass Range & $R$ Mag. Range & 
Min. $\rmerg$ & \underline{Total Number} & \underline{of Mergers} & 
& \underline{$\zres$} & \\
        &        &            &       &                 & 
Main Branch & All Branches & 1:8 & 1:4 & 1:2\\
\hline
\hline
High    &  166  &  $6.4 - 60  \times 10^{10} \Msun$ & 
$-18.3 < M_R < -19.4$  & 0.125 (1:8) & 466 & 595 & 0.5 & 1.0 & 2.6\\
Medium  &  167  &  $3.3 - 6.1 \times 10^{10} \Msun$ &  
$-17.6 < M_R < -18.3$ & 0.250 (1:4) & 132  & 137 & --- & 0.6 &1.0,\\
Low     &  165  &  $2.0 - 3.3 \times 10^{10} \Msun$ & 
$-16.8 < M_R < -17.6$ & 0.500 (1:2) & 65  &  67  & --- & --- & 0.5\\
All     &  507  &  $2.0 - 290 \times 10^{10} \Msun$ & 
$M_R < -16.8$  & ---   & 752  & 988\\
\hline
\end{tabular}
\caption{
The properties of galaxies in our three mass bins are shown
as well as the properties of the sample as a whole.  ``All'' includes 9 
galaxies more massive then the upper mass cutoff of the high mass sample
that are only used in \S~\ref{ssec:vamr}.
The $R$-band magnitude, corresponding to the same number density
of objects in the \citet{blan:03} luminosity function is given for
each mass bin.
The minimum parent mass ratio for which the sample is complete and the 
total number of resolved mergers both along the main branch and for 
all branches are also given.  For the medium and low mass samples 
essentially all resolved mergers are along the main branch.  In the high 
mass sample about $30\%$ of mergers occur off the main branch.  We will
focus on main branch mergers here as they are most important in 
determining the Hubble type of the $z=0$ galaxy. 
}\label{tab:prop}
\end{center}
\end{table*}

\section{Building Merger Trees in the Simulation}
\label{sec:sims}
The simulation was evolved using PTreeSPH \citep{ddh:97}, a parallel
version of the SPH code of \citet{kwh:96}, for a flat 
$\Omega_m=0.4$ cosmology with $\sigma_8 = 0.8$, a Hubble constant
$H_0 = 100h \kms $Mpc$^{-1}$ with $h=0.65$,  a baryon content 
$\Omega_b = 0.047$, and a spectral index $n=0.93$. 
Our choices of $\sigma_8, H_0, n$, and $\Omega_b$ are reasonably close to 
the latest cosmological parameters estimated from the CMB and large scale 
structure \citep{sper:03,eisen:05}, while our value of $\Omega_m$ is higher
by about $1.5\sigma$. 
The box is $22.22h^{-1}$Mpc on a side, with $128^3$ dark matter particles and 
$128^3$ gas particles.  The limited size of the simulation could have
a significant impact on our statistics---in particular there is only 
one cluster mass halo in our simulation volume, and that halo is
itself anomalously large for a box of this size.
The dark matter particle mass is $7.88 \times 10^8 \Msun$, and the 
gravitational softening is a $5h^{-1}$ 
comoving kpc cubic spline, roughly equivalent 
to a Plummer force softening of $3.5h^{-1}$ comoving kpc. 
The gas particles start off with masses of $1.06 \times 10^8 \Msun$, 
though this changes as star particles are created.  
This simulation has previously been used 
for studying quasar absorption systems \citep{gard:01}, 
the X-ray properties of galaxy groups \citep{dkw:02},
the correlation between Lyman Break Galaxies and Ly$\alpha$ forest
absorption \citep{koll:05}, and the histories and mechanisms of 
gas accretion \citep{mura:02,keres:05}. Because the uncertainty in
the Hubble parameter has been significantly reduced with recent observations
and is now known to be quite close to $h=0.7$, we scale all observable
quantities to the values appropriate for $h=0.7$ and do not quote
explicit $h$ dependences.  (For example, we take the box volume to be
$22.22/0.7 = 31.75$ Mpc on a side.)

We analyze $227$ output time steps between $z=8$ and $z=0$,
spaced to give a time resolution better than 150 Myr.
At each step,
galaxies are identified using the Spline Kernel Interpolative DENMAX 
(SKID)\footnote{We use the implementation of SKID by J. Stadel and
T. Quin, which is publicly available at 
http://www-hpcc.astro.washington.edu/tools/skid.html}
algorithm \citep{gb:94,kwh:96}. This algorithm consists of five basic steps:
(1) the smoothed baryonic density field is determined; (2) baryonic particles
are moved towards higher density along the initial gradient of the baryonic
density field; (3) the initial group is defined to be the set of particles
that aggregate at a particular density peak; (4) initial groups that are 
close together are then linked; (5) particles that do not satisfy a negative 
energy binding criterion relative to the group's center of mass are removed
from the group.  These steps are applied to all star particles and to 
gas particles with temperatures $T < 3 \times 10^4$K and densities
$\rho_{\rm gas}/\bar{\rho}_{\rm gas} > 1000$.  The resulting groups of cold
gas and stars have masses and sizes similar to the luminous regions
of observed galaxies \citep{katz:92,kwh:96}.
The smallest systems reliably identified, as determined by
comparison to higher resolution simulations of smaller volumes,
have $64$ particles, or a mass of $6.8\times10^{9}\Msun$. 
However, for our merger study we
restrict ourselves to galaxies with more than three times this mass, or 
$\mgal > 2\times10^{10}\Msun$ at $z=0$, since we also need to identify
their smaller mass progenitors.  

\begin{figure*} 
\centering 
\vspace{0pts} 
\epsfig{file=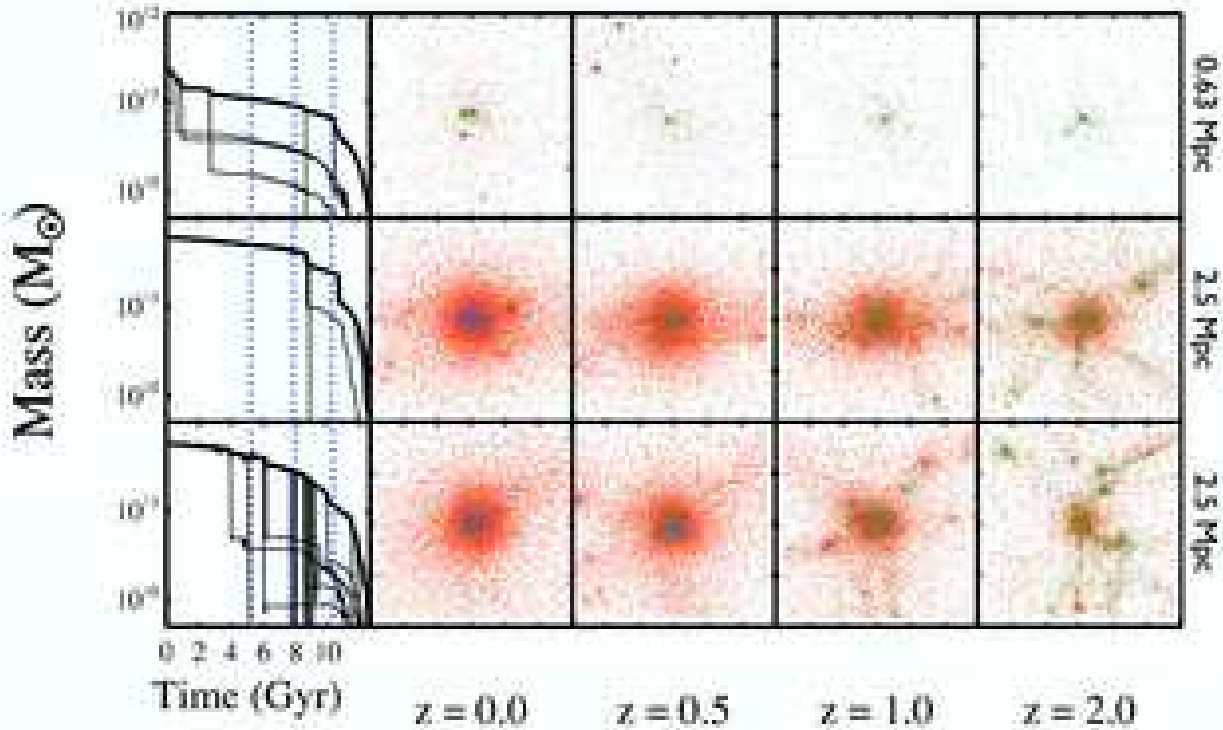,width=\linewidth} 
\vspace{0cm}
\caption{The figure shows the mass accretion histories for three example 
galaxies.  In each row, the leftmost panel shows the masses of all of the
galaxy's progenitors as a function of look backtime.  When galaxies merge the 
lower mass progenitor is connected to the main progenitor by a vertical line.
Vertical dotted lines mark redshifts $z=0.5$, 1, and 2.
The remaining panels show projected particle distributions at $z=0$,
0.5, 1, and 2, in regions 0.63 comoving Mpc (top row) and 2.5 comoving
Mpc (middle and bottom rows) on a side.
Green points denote particles that end up in the galaxy at $z=0$.
Red and blue points mark other gas and star parciles, respectively.
}\label{fig:nar}
\end{figure*}

Previous merger trees that we are aware of have been built by
identifying the progenitors (sometimes only the most massive progenitor) 
of each dark matter halo at time $t_1$ in a previous time step $t_2$, 
then repeating this procedure for all time steps 
\citep{lc:94,wech:02,zhao:03,ljl:03,peir:04,st:04,torm:04}.  However, we have
found a number of complications when trying to apply this algorithm to 
simulated galaxies.  The main difficulty is in reliably defining a galaxy
in each time step.  As is sometimes the case with optical observations,
it can be difficult to determine if merging systems should be identified 
as one or two galaxies, and on rare occasions a galaxy is not identified 
at all in one time step. If a galaxy's 
progenitor is missed in one time step, then the above algorithm will truncate 
the tree and find that a galaxy suddenly appears in the simulation. 
More commonly, we find that two galaxies have the same progenitor, so 
that galaxies appear to split and then re-form into one galaxy (see the
top panel of Figure \ref{fig:nar}).  A third complication is that particles 
can be stripped from both merger progenitors during the merging process but 
later join the merger remnant.  Whether these particles should be counted as 
accreted mass or as part of the merger is again a matter of definition.  
These complications affect only a small minority of galaxies, and
the application of SKID to the star and cold gas distributions generally
yields robust galaxy identifications.  However, with 227 time steps,
even an algorithm that is reliable 99.9\% of the time will encounter
problems for one out of five merger trees.

We have developed a modified merger tree algorithm to address these
issues.  We distinguish between a SKID group, which is the output
from running SKID on the simulation, and a galaxy, which we define in terms
of our merger tree.  Usually the two are identical, but occasionally a galaxy
can be made up of two or more SKID groups, or contain 
particles not in any SKID group, or both.  
We identify all SKID groups at $z=0$, trace each group's members
back through all previous time steps, and identify all progenitor
SKID groups containing at least 16 of these members.\footnote{We
have checked that our trees remain essentially unchanged if we change
this requirement to 8 or 32 particles.}
If a particle is not part of a SKID group in a 
particular time step but is part of a SKID group at an earlier time step,
then we assign it to a ``virtual'' SKID group.  
These virtual assignments allow a SKID group to have multiple descendants,
at least temporarily, usually during and after mergers.  

To build the merger tree of a SKID group identified at $z=0$
(or another redshift chosen for the analysis), we trace its progenitors
back in time but always combine all of the descendants (real and virtual)
of a progenitor SKID group into a single object.  This approach
deals with the problem of temporarily split off particles or divided
galaxies.  We do not count particles that are not in ``real'' SKID
groups at $z=0$.  
Occasionally the two descendants of a single galaxy
never rejoin at a later time.  When this occurs near $z=0$, we interpret 
it as an interrupted merger process and anticipate that the two
descendants would rejoin if the simulation were run into the future.
Therefore, when multiple descendants produce two branches reaching $z=0$,
we combine them into one galaxy if the split occurred within $0.7$ Gyr 
(half the dynamical time of the dark halo).  If the two branches
have remained separate longer than that, we assume that they will never 
rejoin,
and instead separate the common progenitor into two galaxies, each one getting
the descendant's fraction of the total mass of the SKID group.  In these cases
we always find that the SKID group has two parents with masses nearly
identical to those of the two descendants; thus it appears that what has 
happened
is that two galaxies have momentarily passed close enough to one another to 
be identified as one SKID group but do not have a strong physical interaction.
The combining 
of unmerged descendants at $z=0$ means that we start off with 516 SKID groups
that turn into 507 merger trees (galaxies).

As is generally the case in simulations of this sort, efficient cooling
leads to galaxy masses at $z=0$ that are systematically larger
than observational estimates.  Figure~\ref{fig:mf} 
shows the $z=0$ cumulative mass function, $\Phi(\mgal)$, of this simulation 
in comparison to the functional fit to the $z=0$ mass function estimate
of \citet{bmkw:03}, which uses galaxy luminosities and colors to
infer stellar masses assuming a ``diet Salpeter'' initial mass function (IMF).
The cumulative mass function is just the number density
of objects with mass greater than $M$, or 
\beq
\Phi(M) = \int^\infty_M \phi(M') dM'.
\eeq 
The simulated galaxies are too massive at all number densities, with
the discrepancy worsening at low and high mass values.  The source of this
discrepancy remains unclear even after many years of study.  It is 
possible that it is partly a numerical artifact \citep[e.g.][]{sh:02}, or 
the result of insufficient supernova \citep[e.g.][]{ds:86} or AGN 
\citep[e.g.][]{omma:04} feedback in the simulations, or the result
of incorrectly assuming a universal stellar IMF.
\citet{keres:05} argue that a physical mechanism that 
suppresses ``hot'' accretion from shock heated gas halos
would substantially reduce galaxy masses, 
especially above $M_*$.  Such a mechanism might involve AGN feedback
or conduction, or it might simply emerge from a proper treatment of
multi-phase cooling \citep{mb:04}, which can reduce the overall rate
of cooling onto massive galaxies
\citep[see][for a recent review of all these processes]{db:04}.

Because we compute merger statistics in terms of galaxy mass ratios,
we are hopeful that any solution to the global mass discrepancy will have 
only a modest effect on the galaxy merger statistics that we present here.
There may, however, be some changes to merger rates because lower mass
galaxies have longer dynamical friction timescales.  
We plan to investigate the importance of this
effect in future work.  The correction to galaxy masses
needed to make the simulated 
mass function agree with the observations is shown in the bottom panel of 
Figure~\ref{fig:mf}.  In the galaxy mass range that we 
consider, $2\times10^{10}\Msun < \mgal < 6\times10^{11}\Msun$ at $z=0$, 
the correction is well approximated by a constant factor of $2.75$ 
(dotted line).  Thus we 
expect that a physical or numerical fix to the mass discrepancy
would have only modest impact on the merger rates at a given mass ratio.
However, the correction becomes more 
significant at low masses (a factor of 18 for the lowest mass bin), which 
suggests that we may significantly overestimate the number of mergers when
one parent has a mass less than $10^{10}\Msun$.  We will use the galaxy 
masses from the simulations in our discussion, but when connecting to 
observations we note that to get the correct number density of galaxies 
these masses should be divided roughly by a factor of 2.75. 
An alternative way to make the connection to observed galaxy populations
is to match the cumulative space density above mass thresholds in 
our simulation to the cumulative space density computed from the
observed galaxy luminosity function \citep{blan:03}, assuming that
luminosity is a monotonic function of mass.  The luminosities 
implied by this number density matching procedure are given in 
Table~\ref{tab:prop}.

\begin{figure} 
\centering 
\vspace{0pt} 
\epsfig{file=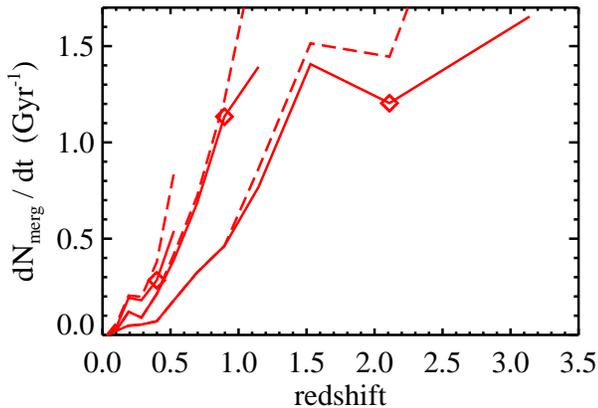,width=\linewidth} 
\vspace{0pt}
\caption{
The number of mergers per galaxy as a function of redshift for the high
mass sample.  The three solid lines are, from top to bottom, 
$\rmerg > 0.125$, $\rmerg > 0.25$, and $\rmerg > 0.5$.  The 
dashed lines show the maximum number of possible mergers including unresolved
mergers.  The diamonds mark the last bin where the Poisson error on the 
resolved mergers is larger than the possible contribution of unresolved 
mergers and the lines are continued for one bin past this point.
The dividing redshift between these bins and the next are 
$\zres = 0.5, 1.0$ and $2.6$. 
}\label{fig:resh}
\end{figure}

\section{The properties of galaxy merger trees}
\label{sec:prop} 
Figure \ref{fig:nar} illustrates the merger history of three simulated
galaxies, selected to show a range of behaviors.
In the left panels, lines show the baryonic masses
(stars plus cold gas) of each galaxy's progenitors as a 
function of lookback time.  In the second column, green points
mark particles that are members of the galaxy at $z=0$, and 
blue points mark stars in other galaxies.  Red points mark the
surrounding gas particles, most of which are at high temperature.
The remaining panels show the positions of these particles at
$z=0.5$, 1, and 2, in regions of constant comoving size.
Green points always represent particles that will be in 
the galaxy at $z=0$.

The topmost galaxy has three clearly identifiable progenitors 
at $z=0.5$.  It experiences two major mergers (roughly 1:3) in 
the final Gyr, and it still has a bimodal appearance at $z=0$.
In fact, SKID splits the system into two components at $z=0$,
but they are a single component at an earlier output and are
therefore counted as a single galaxy.
The galaxy in the middle panels undergoes its last major
merger at $z\sim 1.2$, and thereafter it grows by
smooth accretion, doubling its mass by $z=0$.
The morphological type recipe that we adopt in \S\ref{sec:btd}
(similar to that in semi-analytic models) assigns the post-merger accretion
to a disk component and therefore assigns this galaxy a
bulge-to-total mass ratio of $\sim 0.5$.  However, if some
mechanism suppresses hot gas accretion at late times
\citep{binn:04,kkdw:03,db:04,keres:05}, then the bulge fraction
would be higher (see \S\ref{sec:btd} for further discussion).
The galaxy in the bottom panel experiences many mergers, but
all of them are minor; our morphological type recipe would
therefore identify this as a disk-dominated system.

\begin{figure} 
\centering 
\vspace{0pt} 
\epsfig{file=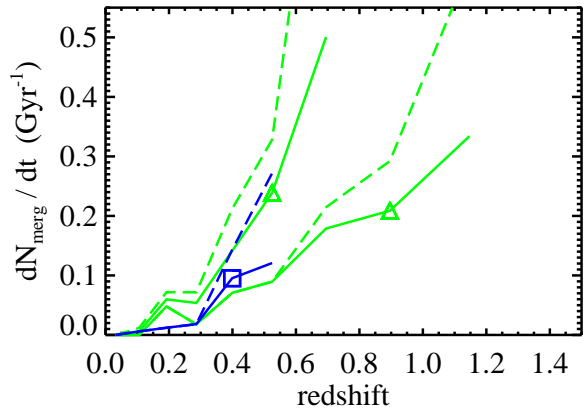,width=\linewidth} 
\vspace{0pt}
\caption{
The number of mergers per galaxy as a function of redshift for the medium
and low mass samples.  The triangles denote the medium mass sample with
the upper solid line being $\rmerg > 0.25$ and the lower solid line
being $\rmerg > 0.5$, while the square marks the line for the low mass
sample with $\rmerg > 0.5$.  The dashed lines show the maximum 
number of possible mergers including unresolved mergers.
The symbols mark the last
bin where the Poisson error on the resolved mergers is larger than the 
possible contribution of unresolved mergers and the lines are continued for 
one bin past this point. The dividing redshift between these bins and the 
next are $\zres = 0.5, 0.6$ and $1.0$ for the low mass sample and the 
medium mass sample with $\rmerg > 0.25$ and $\rmerg > 0.5$, respectively. 
}\label{fig:resl}
\end{figure}

In quantitative terms,
we would like to know the function $\Psi$, where 
\beq
\Psi(\mgal,z,\rmerg) = 
{{\partial P_{merg}}\over{\partial t\,  \partial \mgal\, \partial \rmerg}},
\eeq
the probability that a galaxy with a mass between $\mgal$ and 
$\mgal + d\mgal$ at a redshift $z$ was the result of the merging of two 
galaxies with a mass ratio between $\rmerg$ and $\rmerg + d\rmerg$ 
in the preceding time interval $dt$.  
We define $\rmerg < 1$, i.e., it is the ratio of the baryonic mass of 
the smaller parent to that of the larger parent.
Unfortunately, we do not resolve a large enough dynamic range to fully 
characterize the function $\Psi$, so instead we study several 
interesting ``projections'' of it to learn about its overall parameter 
dependence.

With the morphological assignment recipe of \S\ref{sec:btd},
a galaxy's Hubble type
is determined by the merger history of its ``main'' progenitor, the
most massive progenitor at each output.  
If the Hubble types of the merging galaxies change the 
result of the merger, e.g.\ because bulges behave differently 
from disks, then the merger history of the minor progenitors 
also matters.  However, this effect is
likely to be of secondary importance, and in this paper
we will not distinguish between mergers of disk and bulge galaxies.
Table~\ref{tab:prop} lists the number of resolved main branch mergers
(mergers involving a galaxy's main progenitor) and the total number
of resolved mergers for all of our merger trees.  
Only for the high mass sample, $M_{\rm gal} > 6.4\times10^{10} \Msun$, 
do we resolve a significant number of off-main branch mergers.  
In most cases, therefore, we present results only for
main branch mergers, but we occasionally compare the statistics of
main branch mergers
to those of all mergers for the high mass sample. 

\begin{figure} 
\centering 
\vspace{0pt} 
\epsfig{file=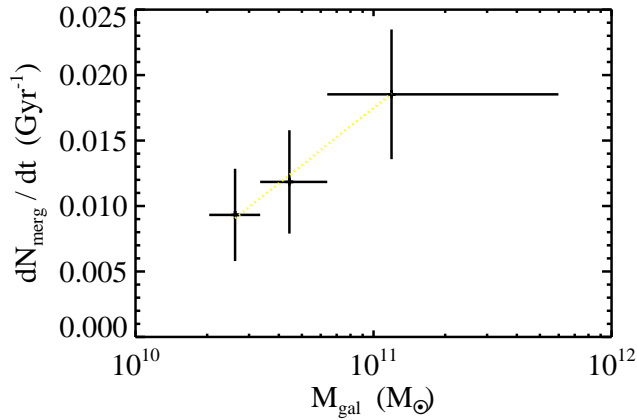,width=\linewidth} 
\vspace{0pt}
\caption{
The number of nearly equal mass mergers ($\rmerg > 0.5$) per galaxy, 
that occur for $0 > z > 0.5$ as a function of the galaxy mass $M_{gal}$ 
at $z=0$.  The number of mergers is strongly dependant on the galaxy 
mass.  The dashed line shows a linear fit to $\log(M_{gal})$ with a 
slope of $0.0145$.
}\label{fig:mass}
\end{figure}

\subsection{Galaxy-averaged merger statistics}

Figures~\ref{fig:resh} and~\ref{fig:resl} present our main
characterizations of $\Psi(\mgal,z,\rmerg)$:
the average number of mergers per galaxy per Gyr above
$\rmerg$ thresholds of 0.125, 0.25, and 0.5 for galaxies in
the high mass sample (Fig.~\ref{fig:resh}) and the medium
and low mass samples (Fig.~\ref{fig:resl}).  
The relation of $dN_{\rm merg}/dt$ to $\Psi(\mgal,z,\rmerg)$ is
\begin{eqnarray}
\frac{dN_{\rm merg}}{dt} &= 
  {1\over N_{\rm gal}} \sum_{i=1}^{N_{\rm gal}}
  \int_{R_{\rm min}}^1 \Psi(M_i, z, R_{\rm merg}) dR_{\rm merg} \\
  &\approx 
  {1\over \bar{n}} \int_{M_{\rm min}}^{M_{\rm max}} dM {dn\over dM} 
  \int_{R_{\rm min}}^1 \Psi(M_i, z, R_{\rm merg}) dR_{\rm merg},
\end{eqnarray}
where $R_{\rm min}$ is the mass ratio threshold, $N_{\rm gal}$ is
the number of simulated galaxies in the mass range $M_{\rm min}$ to
$M_{\rm max}$, $dn/dM$ is the galaxy baryonic mass function, and $\bar{n}$
is the mean space density of galaxies in the mass range.
We compute $dN_{\rm merg}/dt$ by counting mergers in a redshift
interval $\Delta z$ and dividing by the corresponding time interval
$\Delta t$, with typical values $\Delta t \approx 1 {\rm Gyr}$.
The lower mass ratio mergers (i.e., $R_{\rm min}=1/8$ or $1/4$)
can only be resolved for higher mass galaxies.
Merger rates are substantially higher for massive galaxies ---
e.g., the average rate of $\rmerg>0.5$ mergers at $z=0.3$ is
$0.054\,{\rm Gyr}^{-1}$ for the high mass sample and
$0.018\,{\rm Gyr}^{-1}$ for both the medium and low mass samples.

\begin{figure} 
\centering 
\vspace{0pt} 
\epsfig{file=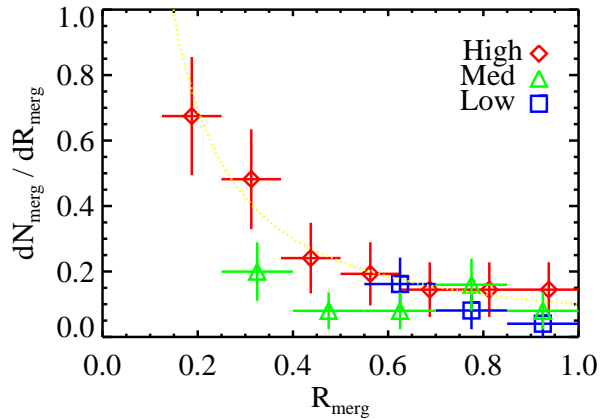,width=\linewidth} 
\vspace{0pt}
\caption{The fraction of mergers with parent mass ratios $\rmerg$ for 
galaxies in the high (diamonds), medium (triangles) and low (square) mass 
samples are shown for mergers in the redshift range $0 < z < 0.5$. The points
are normalized by the width of the bin.  The distribution 
of the high mass sample is well fit by a function $\propto \rmerg^{-1.2}$
plotted as a dotted line.  For the medium and low
mass samples we do not have enough data to constrain the shape of the 
distribution, so we do not know if this distribution is a function of galaxy
mass or redshift.
}\label{fig:ratio}
\end{figure}

Our main limitation in computing these statistics is that we do not 
reliably resolve SKID groups with fewer than 64 particles.  Thus, 
when a galaxy's mass increases by 63 particles we cannot tell without 
detailed examination 
whether this growth was the result of a merger or of smooth accretion. 
The directly calculated rates shown by solid lines in 
Figures~\ref{fig:resh} and~\ref{fig:resl} are therefore lower limits
to the true rates.  Dashed curves accompanying the solid curves
show rates that include the maximum contribution of unresolved mergers,
assigning all growth of up to 63 SPH particle masses that is not
in resolved mergers to unresolved mergers.
At low redshift there are no unresolved mergers,
but as we go back in redshift and the galaxy masses decrease, the number
of possible unresolved mergers increases.  We stop each line when the number
of possible unresolved mergers is greater than the Poisson uncertainty in 
the number of resolved mergers.  Symbols
mark the last output bin before this limiting redshift, 
where we believe we resolve enough of the mergers to robustly
study their properties.
Since the assumption used for the dashed lines is extreme, we expect
that the solid lines are generally closer to the true rates.

For the high mass sample, the limiting redshifts
for $\rmerg > 0.125, 0.25$ and $0.5$ are $\zres = 0.5, 1.0$
and $2.6$, respectively.  
For medium masses, merger ratios of
$\rmerg > 0.25$ and $0.5$ can be resolved to $\zres = 0.6$ and $1.0$,
respectively.  For the low mass sample, 
$\rmerg > 0.5$ can be resolved only to $\zres = 0.5$.  
To compare all three samples, we must therefore restrict ourselves
to $\rmerg>0.5$ and $z \leq 0.5$.  

Figure~\ref{fig:mass} quantifies the mass dependence of merger rates
seen in Figures~\ref{fig:resh} and~\ref{fig:resl}.
To maximize statistics, we count all mergers with $\rmerg > 0.5$ 
occurring between $z=0$ and $z=0.5$. 
The mean number of mergers per galaxy in this interval follows
a best-fit relation
\beq
N_{\rm merg} = 
0.0145 \log_{10}\left({{M_{\rm gal}}\over
{10^{11} \Msun}}\right)+0.0175,
\eeq
where $M_{\rm gal}$ is the galaxy mass at $z=0$.  
We can therefore anticipate a strong dependence of Hubble type
frequencies on galaxy mass.
We find an even steeper best-fit slope for 
$\rmerg > 0.25$, but with only two points it is hard to draw 
meaningful conclusions from this difference.

Figure \ref{fig:ratio} shows the distribution of parent mass ratios for
main branch mergers in the interval $0 < z < 0.5$.  In the high 
mass sample, the number of mergers is proportional to $\rmerg^{-1.2}$.
There are not enough bins in the low and medium mass 
samples to reliably infer a dependence (or lack thereof) on galaxy mass.
The $-1.2$ slope for the high mass sample implies that high mass
ratio mergers dominate the merger growth rate, since the 
average rate at which a galaxy gains mass by mergers above a threshold
$R_{\rm min}$,
\beq
{{dM}\over{dt}} \propto \int_{R_{\rm min}}^1
  \Psi(\rmerg ) \rmerg  d\rmerg   \propto 1-R_{\rm min}^{0.8},
\eeq
has already reached half of its asymptotic value for $R_{\rm min}=0.42$.

\subsection{Volume-averaged Merger Rates}
\label{ssec:vamr}
\begin{figure} 
\centering 
\vspace{0pt} 
\epsfig{file=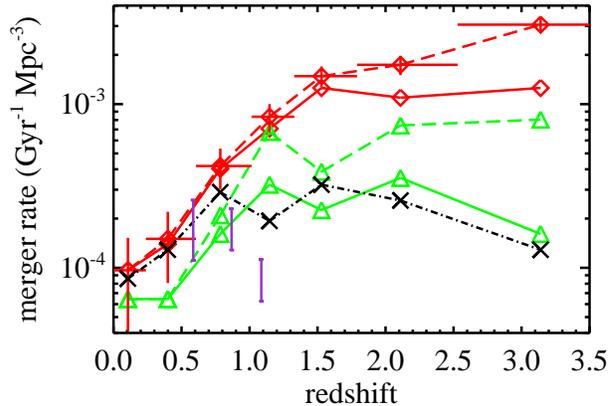,width=\linewidth} 
\vspace{0pt}
\caption{
The number of mergers per Gyr per comoving Mpc$^{3}$ is shown 
as a function of redshift.  The high mass sample (diamonds) and medium mass
sample (triangles) are shown for all resolved mergers (solid lines) and 
including the total possible number of unresolved mergers (dashed lines).  We
see that the contribution of unresolved mergers is negligible for $z < 1.7$
in the high mass sample and $z < 1.0$ in the medium mass sample.  The merger
rate when galaxies are selected by their mass at that redshift is shown as 
the dot-dash line with crosses.  This is more relevant to what one 
sees observationally in a flux limited sample.  For comparison the data of 
\citet{lin:04} is shown as the naked error bars for a similar range in 
stellar mass. The error bars on the upper most points
are representative of the errors for all the simulation points.
}\label{fig:rate}
\end{figure}

We now turn to the volume-averaged 
merger rate of galaxies (the mean number of mergers per 
comoving Mpc$^3$ per Gyr), which can be 
estimated observationally by counting either recent
merger remnants or close pairs that will merge in the
near future, and dividing by an estimated merger timescale.
The merger rate of all galaxies 
(not just the main progenitor branch) is shown in Figure \ref{fig:rate}
for mergers with $\rmerg > 0.5$.
We plot the result for all branches because observations cannot determine
which merger remnants will end up merging with other galaxies
at future times.  Dashed lines include the possible contribution
of unresolved mergers.  We continue curves past the limiting
redshift $z_{\rm res}$ adopted in Figures~\ref{fig:resh} and~\ref{fig:resl}
because 
the range of possible merger rates is still an interesting prediction,
even though the contribution of unresolved mergers may be significant.
Although we are likely to be missing some mergers above $z_{\rm res}$,
we think that the resolved merger rates (solid lines) are probably
closer to the true merger rate predictions than the dashed lines, 
which are upper limits based on extreme assumptions.

The naked error bars in Figure~\ref{fig:rate} show the observational 
estimates of \citet{lin:04} derived from the DEEP2 survey 
\citep{davis:03}, for galaxies with luminosities $-19 < M_B < -21$.  
These points lie below our model predictions, especially at $z>1$.
The luminosity range roughly corresponds to the mass range of our high 
mass sample, except that these are the galaxy luminosities at the 
observed redshift, not at $z=0$. If we restrict ourselves to galaxies 
that are in that mass range at the plotted redshift, then we get the 
dot-dash line with crosses in Figure \ref{fig:rate}, which agrees 
fairly well with the observations at $z<1$ but remains high by a 
factor $\sim 2$ at $z>1$.  Evolution of stellar mass-to-light ratios 
changes the correspondence between luminosity and stellar mass, and 
the range of mass-to-light ratios becomes larger at earlier epochs, so 
a full assessment of this mild discrepancy will require more detailed
modeling of the stellar populations of the simulated galaxies, and
more detailed replication of the observational procedures for
estimating merger rates. 
The systematic uncertainties in matching theoretical and observational
samples are minimized if merger rates are given divided by
the number density of objects being sampled, as in our Figures 
\ref{fig:resh} and \ref{fig:resl}. When data are presented as in Figure
\ref{fig:rate}, the number of mergers per galaxy is convolved with
evolution of the number density of galaxies and of their properties,
making it more difficult to isolate the source of discrepancies.

\begin{figure} 
\centering 
\vspace{0pt} 
\epsfig{file=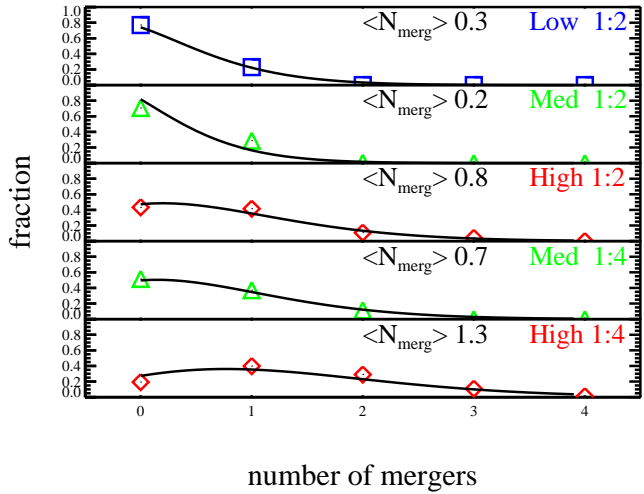,width=\linewidth} 
\vspace{20pt}
\caption{The number of mergers a galaxy undergoes for the low, medium and
high mass samples and for parent mass ratios of $1:2$ and $1:4$.  We see
that all five are well fit by a Poisson distribution with a galaxy having
a mean number of mergers of $0.3, 0.2, 0.8, 0.7$ and $1.3$, respectively. 
}\label{fig:num}
\end{figure}

Figure \ref{fig:num} shows the distribution of the total
number of mergers a galaxy undergoes during
its history, for different mass samples and mass ratio thresholds.  
Figure legends list
the average number of mergers per galaxy for each mass bin and parent 
mass ratio.
Curves in each panel show a Poisson distribution with the corresponding
mean, and in all cases they describe the measured distribution with
reasonable accuracy.  This result suggests that each merger is an
independent event; the average probability of mergers depends on
galaxy mass, but in a given mass range there are not ``merger heavy''
or ``merger light'' galaxies beyond what is expected from Poisson 
statistics.

\subsection{Last Major Merger}

\begin{figure} 
\centering 
\vspace{0pt} 
\epsfig{file=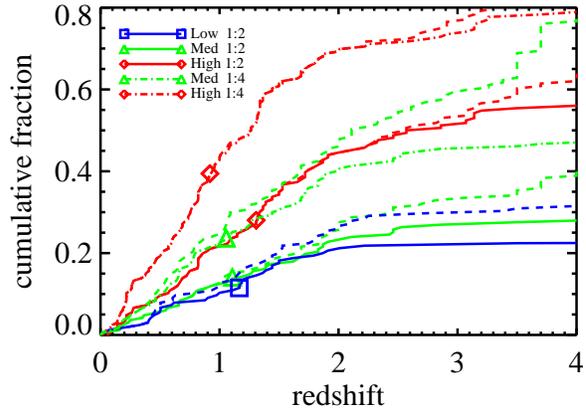,width=\linewidth} 
\vspace{10pt}
\caption{The fraction of galaxies that have undergone a merger of parent
mass ratio $\rmerg > 0.5$ (solid lines) or $\rmerg > 0.25$ (dot-dash 
lines) as a function of redshift is shown for the three mass samples:
high (diamonds), medium (triangles) and low (squares).  We see that in 
all cases some galaxies never undergo a merger of that parent mass ratio.
The corresponding dashed lines show the cumulative fraction including 
unresolved mergers.  The median redshift of galaxies that do undergo a 
merger is marked by the symbols.
}\label{fig:cum}
\end{figure}

We now turn to another interesting characteristic of merger histories,
the distribution of redshifts at which galaxies experience their last
major merger.
Figure~\ref{fig:cum} plots the fraction of galaxies that have
undergone a merger above a 1:2 or 1:4 mass threshold since redshift $z$,
in the three mass bins.  
Including unresolved mergers has only a modest effect on this 
result and is most significant at higher redshifts (see dashed lines).
We see that in all cases a sizeable fraction of galaxies have never undergone 
a merger since the time the galaxy first crossed our mass resolution 
threshold of $6.8 \times 10^9 \Msun$.  
(Galaxies in the low mass sample have grown by a factor of $3-5$ since
that time, while galaxies in the high mass sample have grown by at least
a factor of nine.)
In the medium mass sample, for example, only half of the galaxies
experienced a merger with mass ratio larger than 1:4, and less than
30\% had a 1:2 merger.

The symbols in Figure~\ref{fig:cum} mark the median redshift of the most 
recent major merger for those galaxies that {\it did} experience
a merger.  About half of these mergers occur at $z \la 1-1.5$.
We can compare this result to the predictions for dark matter halos.
\citet{wech:01}, Fig. 137, shows the redshift distribution of the last 
1:3 merger for dark matter halos in differing mass ranges.  If we simply 
approximate the mass of the halo for each of the $z=0$ galaxies by 
multiplying by $2 \Omega_m / \Omega_b$, we obtain
median halo masses of $0.6, 1$ and $2 \times 10^{12} \Msun$ for the low, 
medium and high mass galaxy samples, respectively.\footnote{For 
this halo mass range, we find in the simulation that 
roughly $50\%$ of the halo baryons end up in the central galaxy.}
\citet{wech:01} finds that 40\% and 28\% of halos with masses
of 0.6 and $1.4 \times 10^{12} \Msun$,
respectively, have not had a major merger (greater than 1:3) since $z=4$.
For those halos that did have major mergers at $z<4$, the median
redshift of the most recent merger is $z=1$ in both cases.
Compared to the dark halos, more galaxies in our simulation are merger 
free, and the median redshift of the last major merger is slightly higher.
However, \citet{wech:01} uses a slightly different cosmological
model ($\Omega_m = 0.3$, $\Lambda=0.7$, $\sigma_8=1$, and $n=1$), which
could plausibly account for differences of this magnitude.

The similarity of our galaxy merger rates to the \citet{wech:01}
halo merger rates is broadly consistent with the idea
that most major mergers of halos are quickly followed by
mergers of their central galaxies.  In future work
we intend to perform a more detailed analysis of this point
using halos identified in our simulation to 
investigate the relationship between dark halo and galaxy mergers.
Unfortunately, a comparison of our results to 
semi-analytic models of galaxy formation is 
difficult, since most authors only present results that are meant to be 
compared directly with
observations and do not present the intermediate results necessary
for a comparison to 
the simulations or to other semi-analytic models.  
The only semi-analytic prediction with which we can compare is that of 
\citet{baugh:96}, who show plots similar to Figure~\ref{fig:cum}.
In their model, nearly all galaxies have had a major merger
since $z \sim 2$.  However, they consider an $\Omega_m = 1$ cosmology, 
which is undoubtedly the major cause of this large difference from
our results. 
A comparison of Figure \ref{fig:cum} to semi-analytic predictions 
using the same cosmology would be interesting.

\section{The Distribution of Hubble Types}
\label{sec:btd}
We now turn to determining a galaxy's Hubble type from its merger history.
Simulations have shown that collisions between disk galaxies typically result 
in spheroidal systems \citep{tt:72,barn:88,hern:92,bh:96,nbh:99,nb:03}.
We investigate a simple ansatz for the 
effect of mergers on morphology: if the mass ratio $\rmerg$
exceeds a threshold $\rmaj$, we
convert the entire baryonic mass of the merger remnant to a 
spheroid.\footnote{We have also investigated the recipe used in some
semi-analytic models where for each merger of mass ratio $\rmerg > \rmaj$
twice the mass of the less massive progenitor is converted into bulge, 
and this is done for all mergers along the galaxy's main branch.  
Final bulge-to-disk ratios for this ``partial conversion'' recipe
are slightly lower than those for the ``total conversion'' recipe,
but the qualitative results are very similar, so we do not 
present them separately.}
Mergers below the $\rmaj$ threshold have no effect.
All mass accreted after the last major merger is assigned to 
the disk component.

\begin{figure} 
\centering 
\vspace{0pt} 
\epsfig{file=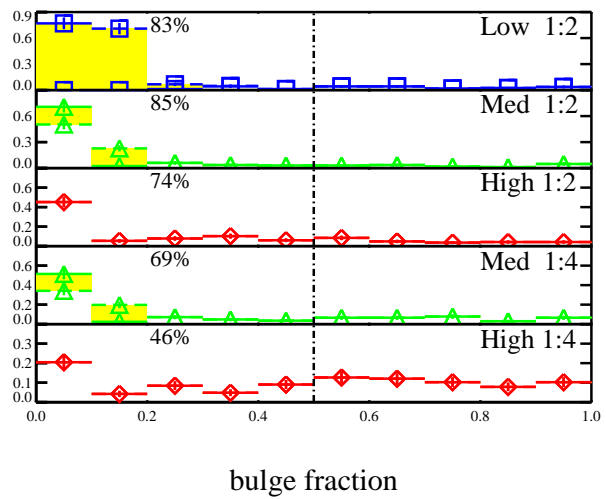,width=\linewidth} 
\vspace{20pt}
\caption{The bulge-to-total fraction is shown for each mass sample 
and for two choices of what constitutes a major merger, $\rmaj = 0.5$ and 
$\rmaj  = 0.25$.  The solid line includes only 
resolved mergers and the dashed line includes all possible mergers. 
The shaded region shows the range that could exist given our finite resolution.
Dividing early and late type galaxies at a bulge-to-total ratio of $0.5$,
we also indicate the late galaxy fraction, which is the same whether or not
unresolved mergers are included.
}\label{fig:bfrac}
\end{figure}

Figure~\ref{fig:bfrac} shows the distribution of bulge-to-total mass
ratios computed using this recipe and the simulated merger histories,
for the three mass bins and $\rmaj$ thresholds of 1:2 and 1:4.
At low bulge-to-total ratio, unresolved mergers could have a significant
impact. At the low end of our mass range we only resolve the later two
thirds of a galaxies mass accretion history, so up to one-third of its
mass {\it could} have had an unresolved merger before we start resolving it.
We therefore show, in addition to our predictions using resolved mergers,
a prediction using the extreme assumption that each galaxy has the
maximum possible contribution from unresolved mergers.  In general,
this maximal contribution simply shifts galaxies from the lowest
bulge-to-total bin ($<0.1$) to the next bin ($0.1-0.2$).
Thus, our prediction for the fraction of truly bulgeless low mass
galaxies can be significantly affected by unresolved mergers, but
our other predictions cannot.

For bulge-to-total ratios $> 0.2$, the predicted distributions in
Figure~\ref{fig:bfrac} are strikingly flat.
In particular, the simulation produces
relatively few systems that are close to 100\% bulge, because after mergers 
galaxies continue to accrete gas and form new stars in a disk component
in the simulation.  
As discussed by \citeauthor{keres:05} (\citeyear{keres:05}; see also
\citealt{binn:04,crot:05,db:04}), 
a mechanism that halts late-time
accretion onto high mass galaxies, such as AGN feedback 
\citep{bt:95} or multi-phase cooling \citep{mb:04}
would improve the agreement between the
simulations and observations by predicting lower masses and older
stellar populations at the high end of the galaxy luminosity function.
Reproducing the correct distribution of bulge-to-disk ratios could
be a powerful test of such mechanisms, since the fraction of 
bulge-dominated systems will change depending on the conditions
assumed for shutoff of gas accretion.

As expected, the higher merger rates of massive galaxies lead to a greater
frequency of high bulge-to-disk ratios in our highest mass bin.
To summarize the mass dependence and compare to observations, we separate
the galaxy populations into ``early'' and ``late'' subsets at a bulge
mass fraction of 0.5.  For a major merger threshold of 1:2, the late-type
fraction is about 85\% in the low and intermediate mass bins but only 74\%
in the high mass bin.  For a threshold of 1:4, the late-type fraction is
69\% in the intermediate mass bin and 46\% in the high mass bin; we
cannot resolve 1:4 mergers in the low mass bin.

\begin{figure} 
\centering 
\vspace{0pt} 
\epsfig{file=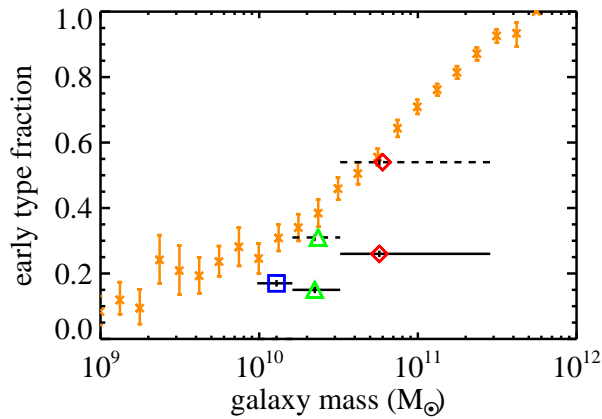,width=\linewidth} 
\vspace{0pt}
\caption{The fraction of galaxies that are early type as a function of the
galaxies mass is shown for parent mass ratios of 1:2 (solid line) and 1:4
(dashed line). The crosses with error bars are the observed fraction 
derived from \citet{bmkw:03}.
The masses of the simulated galaxies have been divided by a factor of 2.75
to bring them into agreement with the observed stellar mass function.
}\label{fig:mdepend}
\end{figure}

Figure~\ref{fig:mdepend} compares these results to an estimate of 
the early-type fraction derived from Bell et al.'s (\citeyear{bmkw:03}) 
determination of the early and late type stellar mass functions, 
using the 2MASS data set \citep{skrut:97,mmkw:05}.  
Because of the offset between predicted and observed galaxy mass
functions discussed in \S\ref{sec:sims}, we divide the simulated galaxy
masses by 2.75 for this comparison; we are thus comparing populations
of similar space density.  For $R_{\rm major}=$1:2, the predicted
early-type fraction is substantially below the observational estimate.
For $R_{\rm major}=$1:4, the agreement is excellent, albeit with only
two theoretical data points.  We have checked that ratios of 1:3 and 1:5
are also in fairly good agreement.  \citet{bmkw:03} determine galaxy
types from concentrations, and the principal systematic uncertainty in
this comparison is that we do not know how well their definition
corresponds to our theoretical criterion $f_{\rm bulge}=0.5$.
The full distribution of bulge-to-disk ratios would be a powerful
test of our predictions and, more generally, a
powerful diagnostic of the physical processes that produce galaxy bulges.
Extracting unbiased estimates of this distribution for large, well defined
galaxy samples is a challenging task, but within reach of modern data
sets \citep[e.g.][]{mmkw:02}.

\section{Conclusions}
\label{sec:conc}

We have used an SPH simulation of a $\Lambda$CDM universe to calculate
the statistics of galaxy merger histories.  Our principal findings
are as follows:

1. The incidence of major mergers is much higher for high mass
galaxies.  At $z=0.3$, we find a rate of 0.054 mergers per Gyr above a 
1:2 mass threshold for galaxies with $\mgal>6.4\times 10^{10}\Msun$,
but the rate for our lower mass galaxy samples is three times lower.

2. The distribution of merger mass ratios is $N\propto \rmerg^{-1.2}$
in our high mass sample (the only one for which our statistics are
sufficient to make the estimate).  With this logarithmic slope,
major mergers dominate the total mass growth from mergers, with half of
the mass coming in mergers above a mass ratio threshold $\rmerg=0.4$.
However, as shown by \cite{mura:02} and \cite{keres:05}, even the
high mass galaxies in our simulations gain most of their mass
through smooth gas accretion rather than mergers with pre-existing
galaxies.  Our derived mass ratio distribution reinforces their
arguments that the dominance of smooth accretion over mergers is
not an artifact of limited numerical resolution.

3. For each mass and mass ratio bin, the merger rate climbs
rapidly with redshift, roughly doubling between $z\sim 0$ and $z\sim 0.6$.
The predicted merger rates are in reasonable agreement with 
observational estimates, but the systematic uncertainties in the
comparison are large because of the difficulty of matching simulated
and observed galaxy populations.

4. In each bin of mass and mass ratio, the number of mergers
per galaxy is Poisson distributed with respect to the mean number
for the bin.  This suggests that each merger is, effectively,
an independent event.

5. A substantial fraction of galaxies experience no major
mergers after $z=1$, and many have no major mergers
over the entire time that they are resolved in our simulations.
For example, even in our high mass sample, only 45\% of galaxies
have an $\rmerg > $1:4 merger at $z<1$, and only 20\% have an
$\rmerg > $1:2 merger.  For those galaxies that do experience major
mergers, the median redshift of the most recent event is $z\sim 1-1.5$
in all mass and $\rmerg$ bins.  

The high frequency of quiescent merger histories is perhaps our
most important result.  If major mergers are indeed the primary
mechanism of spheroid formation, then this large population of 
merger-free galaxies shows that the $\Lambda$CDM model can
easily produce an acceptable fraction of late-type, nearly
bulgeless systems.  The predominance of quiescent merger histories
is also good news for the galaxy angular momentum problem, since
low redshift mergers are frequently responsible for removing
angular momentum and shrinking disk sizes in simulations
that focus on the formation of individual galaxies.

We have combined our merger histories with a simple,
semi-analytic style recipe for spheroid formation to estimate
bulge fractions of our simulated galaxies.
With a bulge conversion threshold of 1:4, we find acceptable
agreement with the observed trend of early-type galaxy fraction with 
galaxy mass, once we rescale the simulated galaxy masses to match the 
observed galaxy mass function.
However, since most galaxies have continuing gas accretion after
their last major merger, the fraction of truly bulge-dominated
systems (i.e., $f_{\rm bulge} \ga 0.8$) is small, even at high masses.
This result suggests that some mechanism that shuts off gas accretion
in high mass halos is needed to explain the observed distribution
of galaxy morphologies.  Many authors have already argued that
a mechanism of this sort is required to reproduce the observed exponential
cutoff of the galaxy luminosity function and the red colors of
the most massive galaxies 
(e.g., \cite{kwg:93,bens:03,binn:04,crot:05,db:04,keres:05}).

Our simulation volume is too small to allow a detailed investigation
of the morphology-density relation.  However, in the one cluster
mass halo ($M_{\rm vir} = 3\times 10^{14}\Msun$) halo that forms
in our simulation, we do not find the high fraction of early-type
galaxies seen in observed clusters.  This discrepancy suggests
that some mechanism other than major mergers must contribute
to morphological transformations in the cluster environment,
such as ram pressure stripping \citep{gg:72} or 
``harassment'' by weak perturbations \citep{moore:96}.

We hope to extend this work in future studies that use
simulations of larger dynamic range.  These will allow us to
investigate the correlation of merger histories with 
environment and to better characterize the merger histories
of lower mass galaxies.  We will also investigate the 
relationship between individual galaxy and dark halo merging histories,
and we will check whether mechanisms that reduce galaxy
baryon masses, such as galactic winds and suppression of hot
accretion, also change the statistics of galaxy mergers.
Comparison of our results to observations already suggests that 
the simplest picture of spheroid formation by mergers is not
the full story.  Future comparisons drawing on better simulations
and more detailed observational analyses should yield much greater
insight into the physics that determines galaxy morphologies.

\section*{Acknowledgments}
We thank Risa Wechsler and Mark Fardal for comments on an earlier draft of 
this paper. This project was supported by NASA ATP grant NAG5-13308, 
NASA LTSA grant NAG5-13102, and NSF grant AST-0205969.

\bibliographystyle{mn2e}         
 
\bibliography{me,gf,hydro,xray,dm,abs,cosmo,spin,coll,2mass,gals} 

\end{document}